\documentclass[aps,twocolumn]{revtex4-1}
\usepackage{graphicx}
\usepackage{hyperref}
\usepackage{amssymb}
\usepackage{dcolumn}
\usepackage{float}
\usepackage{bm}
\usepackage{multirow}
\usepackage{booktabs}
\usepackage[T1]{fontenc}
\usepackage[utf8]{inputenc}
\usepackage{lmodern}
\usepackage{color}
\usepackage{makecell}
\usepackage{tikz}
\usepackage{pgfplots}
\usepackage{mathtools}
\usepackage{amsmath}

\newcommand\Fontvl{\fontsize{8}{5.0}\selectfont}

\newcommand\Fontvm{\fontsize{8}{5.0}\selectfont}

\newcolumntype{M}[1]{>{\centering\arraybackslash}m{#1}}

\DeclareMathAlphabet{\bi}{OML}{cmm}{b}{it}

\def\be{\begin{equation}}
\def\ee{\end{equation}}
\def\bearr{\begin{eqnarray}}
\def\eearr{\end{eqnarray}}

\begin{document}
	
\title{Quantum thermoelectrics based on 2-D Semi-Dirac materials}
\bigskip
\author{Alestin Mawrie}
\email{ales@ee.iitb.ac.in}
\author{Bhaskaran Muralidharan}
\email{bm@ee.iitb.ac.in}
\normalsize
\affiliation{Department of Electrical Engineering, Indian Institute of Technology Bombay, Powai, Mumbai-400076, India}
\date{\today}
\begin{abstract}
We show that a gap parameter can fully describe the merging of Dirac cones in semi-Dirac materials from $K$- and $K^\prime$-points into the common $M$-point in the Brillouin zone. We predict that the gap parameter manifests itself by enhancing the thermoelectric figure of merit $zT$ as the chemical potential crosses the gap followed by a sign change in the Seebeck coefficient around the same point. Subsequently, whenever the chemical potential crosses the gap potential parameter, there is a well-balanced maximum of the power factor and the efficiency of the thermoelectrics. An optimal operating point where co-maximization of the power-efficiency is consequently singled out for the best thermoelectric performance. Our work paves the way for the use of 2D semi-Dirac materials for thermoelectric applications.
\end{abstract}

\maketitle
In the past decade, research on Dirac semimetals has been at the center of condensed matter physics. A Dirac semimetal hosts massless Dirac fermions for which graphene\cite{graphene1,graphene2} has since been a benchmark. Besides the apparently linear dispersion in most Dirac semimetals, some also show additional properties like tilted Dirac cones as seen in organic compounds $\alpha$-(BEDT-TTF)$_2$I$_3$ \cite{tilt1,tilt2,tilt3,tilt4}, 8-\textit{Pmmn} borophene\cite{boro1,boro2,boro3}. Also, in some engineered 2D materials (called semi-Dirac materials), such as TiO$_2$/V$_2$O$_3$ nanostructures\cite{ms1}, 
and dielectric photonic systems\cite{ms4} and hexagonal lattices in the presence of a magnetic field \cite{ms3}, it has been realized that the dispersion exhibits a simultaneous massless Dirac and massive fermion characteristics along two different directions. 
\\
\indent 
\begin{figure}[b]
	\begin{center}
		\includegraphics[width=60mm,height=42.5mm]{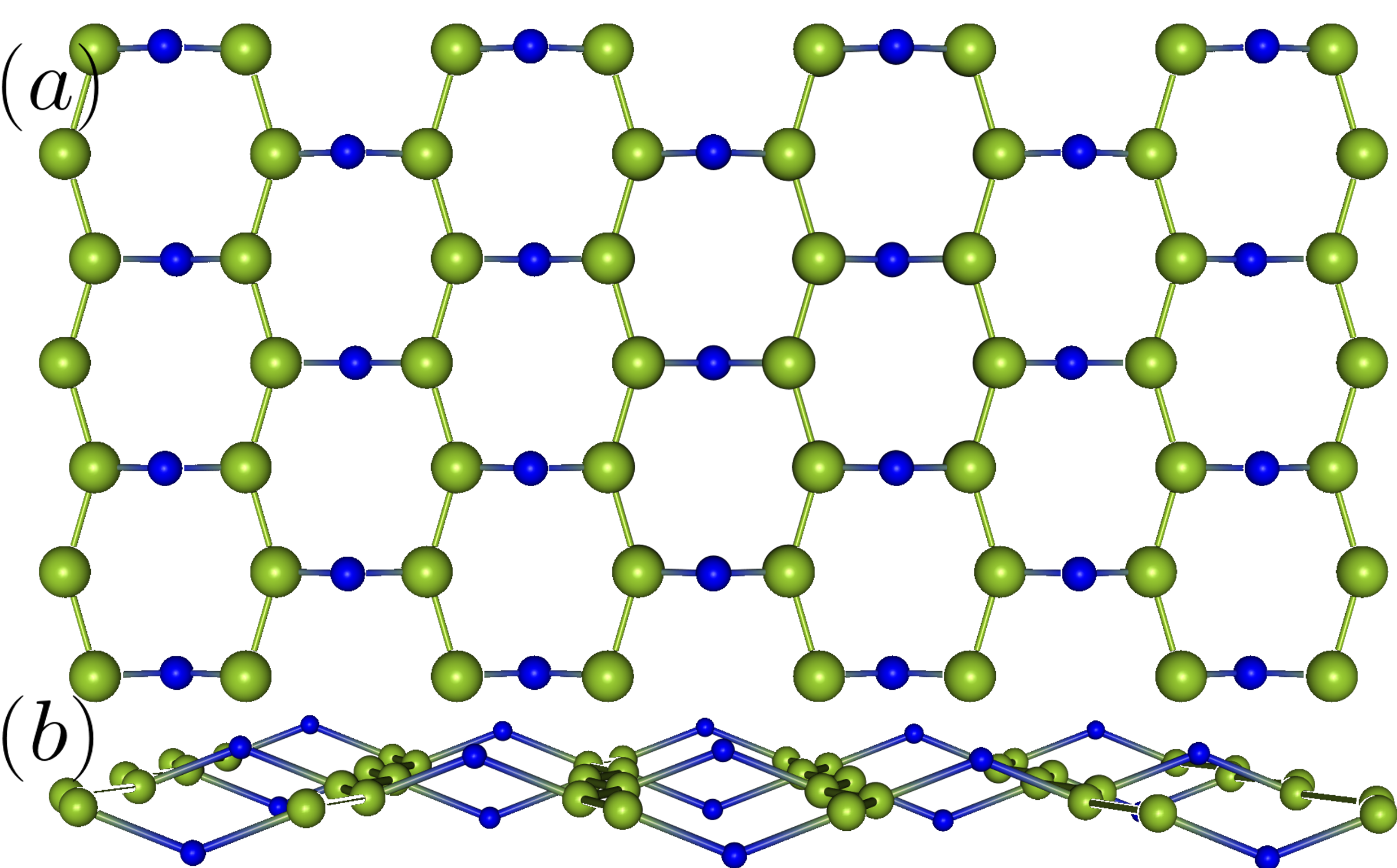}
		\includegraphics[width=24mm,height=42.5mm]{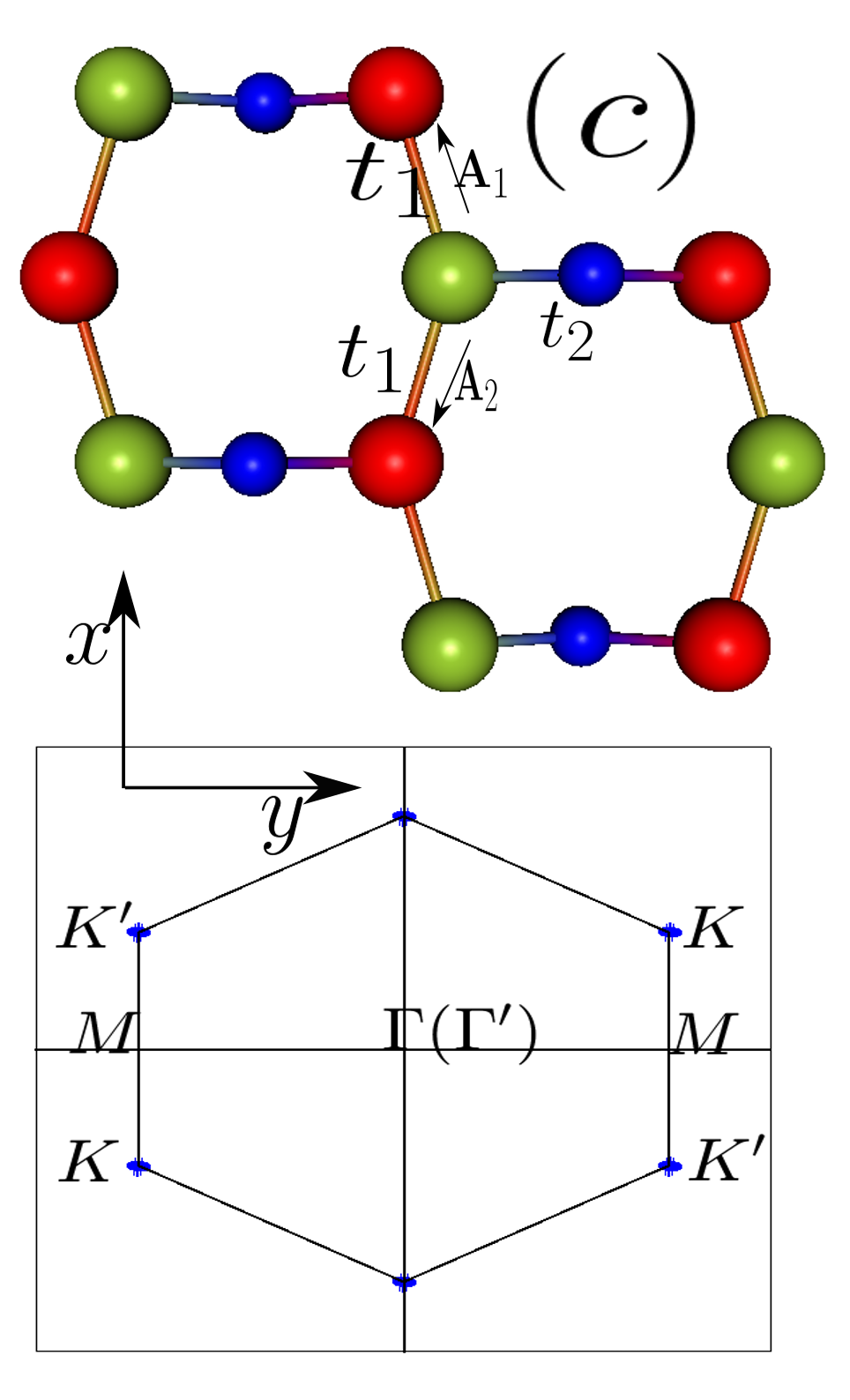}
		\includegraphics[width=55mm,height=55.5mm]{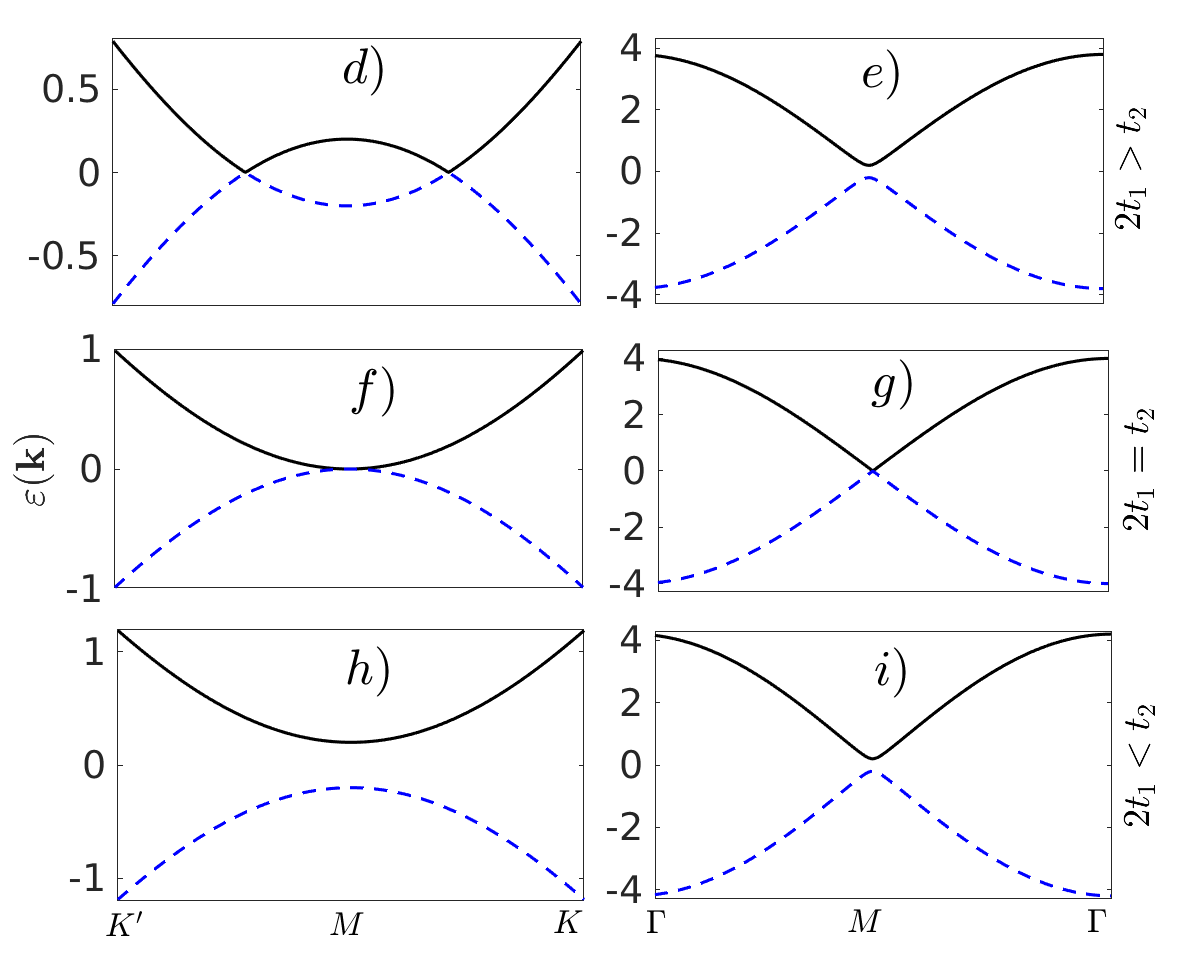}
		\includegraphics[width=25mm,height=55.5mm]{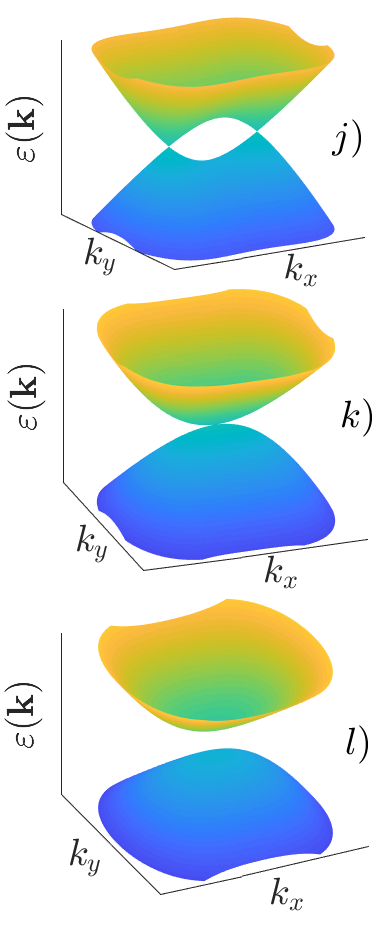}
		\caption{\Fontvm{(a) Top view and (b) side view of the crystal structure of silicene oxide. The green and blue spheres represent silicon and adsorbed oxygen atoms, respectively. (c) The nearest hopping parameters in the honeycomb lattice and the first Brillouin zone with different high-symmetry	points. (d-i) The dispersion along the directions $K\rightarrow M \rightarrow K^\prime$ (d,f,h) and $\Gamma\rightarrow M \rightarrow \Gamma^\prime$ (e,g,i).} (j-l) The low-energy band dispersion for $\delta>0$ (j), $\delta=0$ (k) and $\delta<0$ (l)}\label{Fig1}
	\end{center}
\end{figure}
Recent studies also reveal that a semi-Dirac dispersion can be engineered in some honeycomb based lattices like silicene\cite{Zhong} through a covalent addition of group-VI elements. 
We show that a tight binding Hamiltonian\cite{TbH} that includes different hopping strengths through different types of bonds in oxygen adsorped silicene reveals the merging of the Dirac cones from the $K$- and $K^\prime$-points into a common $M$-point. This further results in the semi-Dirac dispersion around the $M$-point. We later show that this entire process of Dirac cone merging is fully described by a gap parameter. The impact of tunable gap parameter has also been shown to result in a giant and a robust anisotropic optical conductivity in semi-Dirac materials\cite{ales,jang}. Furthermore, a number of interesting properties have also been predicted which attribute to the unique anisotropic dispersion nature of semi-Dirac semimetals. Besides the plausible anisotropic transport properties such as diffusion \cite{diff}, optical conductivity \cite{ales}, semi-Dirac materials also show the appearance of distinct Landau-level spectra under a magnetic field\cite{ms3,Hamilt2}. In this work, we demonstrate how to exploit this unique feature of the merging of Dirac cones to enhance and optimize the thermoelectric device performance featuring semi-Dirac materials.\\
\indent
One of the challenges in designing thermoelectric devices is obtaining optimal conditions that ensure the operation of the thermoelectric nanodevice with maximum power at the best possible efficiency. The maximum efficiency of the system depends on the figure-of merit $zT$ of the system which is defined as $zT=\frac{\mathcal{L}_0 S^2}{\kappa} T$, where $\mathcal{L}_0$ is the electronic conductivity, $S$ is the Seebeck coefficient and $\kappa$ is the thermal conductivity of the system. Attempts have been made to enhance the $zT$-value of several thermoelectric devices by decreasing the thermal conductivity in several alloyed, nanostructured, nanocomposite materials, magnetic graphene and hexagonal boron phosphide bilayer \cite{zT1,zT2,zT3,zT4,Ther,BP}. 
In this letter, using a 2D semi-Dirac based nano-thermoelectric device, we present results that suggest the possibility to obtain optimal power and efficiency besides the enhancement of the thermoelectric parameters. Such a well-balanced maximization feature between the power and efficiency has also been reported in interacting quantum dot thermoelectric setups \cite{bm,bd}. 

%

We begin by understanding how to engineer a semi-Dirac dispersion in certain honeycomb lattices by oxidizing or chemically adsorbing them with other atoms. This is sustained by the possibility that some honeycomb based lattices such as silicene,
germanene and stanene can be easily oxidized or chemically adsorbed by virtue of their buckled structure\cite{ox1,ox2,ox3,ox4,ox5}. In the case of silicene, a covalent addition of a
group-VI element such as oxygen
results in a silicene oxide with the chemical formula of Si$_2$O. 
The crystal structure of Si$_2$O is shown in Fig. \ref{Fig1}, with the top view \ref{Fig1}(a) and the side view \ref{Fig1}(b). The blue and
green spheres represent the silicon and adsorbed oxygen atoms respectively. One way to realize the semi-Dirac dispersion in such a system is by adopting the tight binding Hamiltonian\cite{TbH} as given below
\begin{eqnarray}\label{TBH}
H=\sum_{\bf P} \big[&&t_2\hat{C}^\dagger_{B,{\bf P}}\hat{C}_{A,{\bf P}}+t_1\hat{C}^\dagger_{B,{\bf P}}\hat{C}_{A,{\bf P}+{\bf A}_1}\nonumber\\&&+t_1\hat{C}^\dagger_{B,{\bf P}}\hat{C}_{A,{\bf P}+{\bf A}_2}\big],
\end{eqnarray}
where $t_1$ is the hopping parameter between the non-oxygen adsorped Si-Si bond (\textit{i. e.} in between the Si atoms at $(0,0)$ and at $(\pm\frac{\sqrt{3}}{2},-\frac{1}{2})$; setting the Si-Si bond distance to be $a=1$), and $t_2$ is the hopping parameter between the oxygen adsorped Si-Si bond (\textit{i. e.} between the Si atoms at $(0,0)$ and at $(0,1)$), and $\hat{C}^\dagger_{A/B,{\bf P}}/\hat{C}_{A/B,{\bf P}}$ are the creation/annihilation operators at the site $A/B $ given by the green/red sphere in Fig. [\ref{Fig1}(c)]. The enhancement in the hopping parameter $t_2$ is associated with the change in Si-Si bond length after oxygen adsorption followed by the slight weakening of the corresponding hopping parameter $t_1$ corresponding to the Si-Si bond\cite{Zhong}.
\\
\indent
The dispersion relation corresponds to the Hamiltonian in Eq. (\ref{TBH}) is 
{\Fontvl\begin{eqnarray}\label{kspace}
\varepsilon_\lambda({\bf k})=\lambda\sqrt{2t_1^2+t_2^2+2t_1^2\cos\sqrt{3}k_x+4t_1t_2\cos\frac{3}{2}k_y\cos\frac{\sqrt{3}}{2}k_x},
\end{eqnarray}}
where $\lambda=+/-$ represents the conduction/valence band. The Hamiltonian in Eq. \ref{TBH} helps in drawing a guideline of understanding of the merging of Dirac cones. This is demonstrated in Fig. \ref{Fig1} (d-g). The plot of the dispersion relation in Eq. \ref{kspace} for different ratios of the hopping parameter $t_1$ and $t_2$ is shown in Figs. \ref{Fig1} (d-i). The condition $t_1=t_2$, is an obvious case of graphene where the Dirac nodes appear at all $K$ and $K^\prime$ points in the Brillouin zone. With increasing strength of the parameter $t_2$, the two Dirac nodes at $K$ and $K^\prime$ move closer till they merge at the $M$ point when $t_2=2t_1$ [Figs. \ref{Fig1} (f \& g)]. The experimental realization for the merging of Dirac cones has been
observed in optical lattices\cite{Optical_lattice}. When $t_2$ is slightly less than $2t_1$, [Figs. \ref{Fig1} (d \& e)], the dispersion is gapless at two points along $K^\prime\rightarrow M\rightarrow K$ direction and are gapped elsewhere.
The condition $t_2>2t_1$ represents a gapped dispersion everywhere in the Brillouin zone as seen in Figs. \ref{Fig1} (h \& i) which are of interest here. 
From Eq. \ref{kspace} and also as evident in Figs. \ref{Fig1} (d-i), the dispersion is massive in the direction $K^\prime\rightarrow M\rightarrow K$ and massless along $\Gamma^\prime\rightarrow M\rightarrow \Gamma$. Thus the model provided by the Hamiltonian in Eq. \ref{TBH} garners a full description of the nature of dispersion in semi-Dirac materials. \\
\indent
In the proximity of the $M$ point, the Hamiltonian can be written as
\begin{eqnarray}\label{Hamil}
H={\bf g}\cdot\boldsymbol{\sigma},
\end{eqnarray}
where ${\bf g}=(\alpha k_x^2-\delta,vk_y,0)$ and $\boldsymbol{\sigma}=(\sigma_x,\sigma_y,\sigma_z)$ are the Pauli's spin matrices, with $\alpha$, $\delta$ and $v$ representing the inverse of the quasiparticle mass along the $x$-direction, the system gap parameter and the Dirac quasiparticle velocity along the $y$-direction, respectively. The information on the hopping strength between the corresponding nearest atoms is well-contained in the gap parameter $\delta$. It is zero for $t_2=2t_1$, positive for $t_2<2t_1$ and negative for $t_2>2t_1$. A side-by-side demonstration of the dispersion as given by Eq. (\ref{Hamil}) comparing with that given in Eq. (\ref{TBH}) is shown in Fig. \ref{Fig1} (j-l) and Fig. \ref{Fig1} (d-i), where the conditions of $\delta>0$, $\delta=0$ and $\delta<0$ is demonstrated in Fig. \ref{Fig1} (j), \ref{Fig1} (k) and \ref{Fig1} (l), respectively. For $\delta>0$, there is a van Hove singularity in the density of states when the energy is around $\delta$ \cite{ales}. 
\\
\indent

\begin{figure}[t]
	\includegraphics[width=60mm,height=40.5mm]{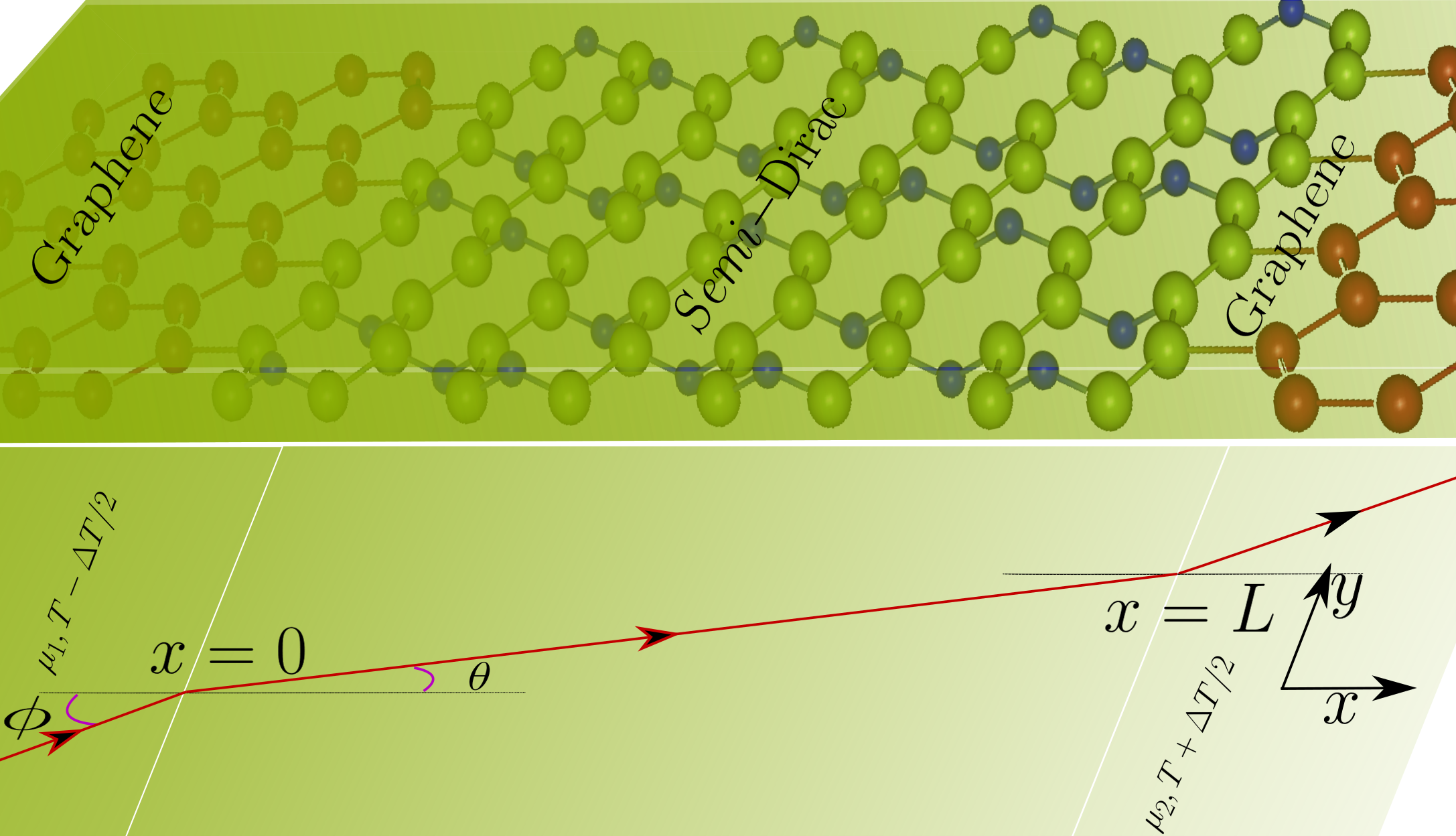}
	\caption{Schematic illustration of a semi-Dirac layer sandwiched between two graphene layers at $x=0$ and $x=L$, respectively.}
	\label{Fig2}
\end{figure}
Having demonstrated the merging of Dirac cones, we now propose a quantum thermoelectric nanosystem based on semi-Dirac materials as shown in Fig. \ref{Fig2}. We consider the 2D semi-Dirac materials sandwiched in between the right and left graphene leads maintained at different chemical potentials and temperatures. The thermoelectric system as demonstrated in Fig. \ref{Fig2} is maintained at potential difference, $(\mu_2-\mu_1)/e$ with temperature gradient $\Delta T$. The electrical and thermal currents in terms of the various Onsager coefficients are as follows
\begin{eqnarray}
\begin{pmatrix}
j\\j_q
\end{pmatrix}=\begin{pmatrix}
\mathcal{L}^{11} & \mathcal{L}^{12}\\
\mathcal{L}^{21} & \mathcal{L}^{22}
\end{pmatrix}
\begin{pmatrix}
(\mu_2-\mu_1)/e\\\Delta T
\end{pmatrix}.
\end{eqnarray}
The different Onsager coefficients are given as 
\begin{eqnarray}
\begin{pmatrix}
\mathcal{L}^{11} & \mathcal{L}^{12}\\
\mathcal{L}^{21} & \mathcal{L}^{22}
\end{pmatrix}=\begin{pmatrix}
\mathcal{L}^{0} & \mathcal{L}^{1}/eT\\
\mathcal{L}^{1}/e & \mathcal{L}^{2}/e^2T
\end{pmatrix},
\end{eqnarray}
where in the ballistic transport regime, within the Landauer-B\"uttiker approach, the coefficients $\mathcal{L}^\alpha$ can be written as
\cite{Ther}
{\Fontvl\begin{eqnarray}
\mathcal{L}^\alpha=G_0\int_{-\pi/2}^{\pi/2}d\phi\cos\phi\int_{-\infty}^\infty d\varepsilon(\varepsilon-\mu)^\alpha\big(-\frac{\partial f}{\partial \varepsilon}\big)\rho(\varepsilon)\mathcal{T}(\varepsilon,\phi).
\end{eqnarray}}
\begin{figure}[t]
	\includegraphics[width=91.5mm,height=60mm]{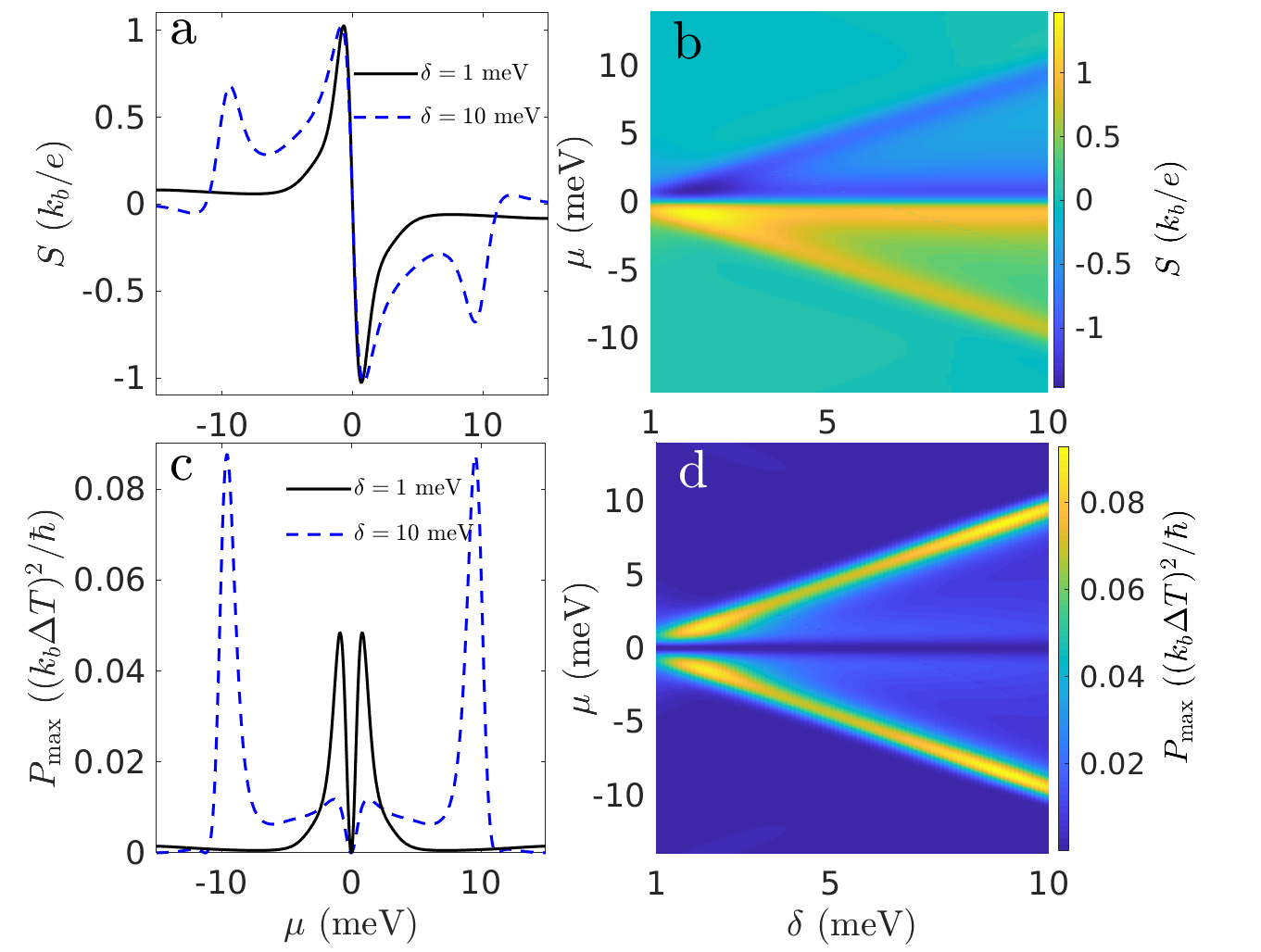}
	\caption{(a \& c) Plots of the Seebeck coefficient $S$ (in units of $k_B/e$), maximum power (in units of $[(k_B\Delta T )^2/h]$ ) for $\delta=1$ meV (solid black curve) and for $\delta=10$ meV (dashed blue curve). (b \& c) Color plots of the Seebeck coefficient and maximum power as a function of chemical potential and the gap parameter.}
	\label{Fig3}
\end{figure}
Here, $G_0=(e^2/\hbar)(W/\pi^2)$, $\rho(\varepsilon)$ the density of states in the semi-Dirac region and $\mathcal{T}(\varepsilon,\phi)$ is the transmission probability of the electron of energy $\varepsilon$, incident at an angle $\phi$. Our ballistic transport assumption remains valid as in the case of 2D materials such as high-mobility suspended or encapsulated graphene\cite{Cala,Bans,wangL}. In order to calculate the transmission probability, we equate the wavefunctions at the boundary $x=0$ and $x=L$. To begin with, the wave functions for
the three regions (as demonstrated in Fig. \ref{Fig2}) for $A$ and $B$ sublattices are:
\\
for $x < 0$,
\begin{eqnarray}
\begin{pmatrix}
\psi_A^I(x,y)\\
\psi_B^I(x,y)
\end{pmatrix}=\begin{pmatrix}
e^{i k_{1x} x}+re^{-i k_{1x} x}\\
e^{i k_{1x} x+i \phi}-re^{-i k_{1x} x-i\phi}
\end{pmatrix}e^{i k_y y}
\end{eqnarray}
for $x > 0$ and $x <L$,
\begin{eqnarray}
\begin{pmatrix}
\psi_A^{II}(x,y)\\
\psi_B^{II}(x,y)
\end{pmatrix}=\begin{pmatrix}
p e^{i k_{x} x}+q e^{-i k_{x} x}\\
p e^{i k_{x} x+i \theta}-q e^{-i k_{x} x-i\theta}
\end{pmatrix}e^{i k_y y}
\end{eqnarray}
for $x >L$,
\begin{eqnarray}
\begin{pmatrix}
\psi_A^{III}(x,y)\\
\psi_B^{III}(x,y)
\end{pmatrix}=\begin{pmatrix}
t e^{i k_{3x} x}\\
t e^{i k_{3x} x+i \phi}
\end{pmatrix}e^{i k_y y},
\end{eqnarray}
which leads to the transmission probability as below
\begin{eqnarray}\label{Transmission}
\mathcal{T}(\varepsilon,\phi)=\frac{1}{\cos^2k_{x}+\sin^2k_{x}\frac{(1-\sin\theta\sin\phi)^2}{\cos^2\theta\cos^2\phi}}.
\end{eqnarray}
It is interesting to note that there is maximum transmission probability for a normally incident right moving electron. The value of $k_x$ as a function of energy and angle of incident is taken depending on the nature of the dispersion in region $II$. In the geometry considered in Fig. \ref{Fig2},  $\varepsilon=\lambda\sqrt{(\alpha k_{x}^2-\delta)^2-(vk_{y})^2}$. Taking the dispersive nature on the graphene side to be $\varepsilon={\bf k}$ and considering the geometry of the problem, $k_y$ remains $\varepsilon\cos\phi$, which ultimately leads to $k_{x}=\sqrt{\frac{\delta\pm\sqrt{\varepsilon^2(1-v^2\cos^2\phi)}}{\alpha}}$.  
Finally, the transmission probability is written as 
\begin{widetext}
\begin{eqnarray}\label{Tr_2}
\mathcal{T}(\varepsilon,\phi)=
\frac{1}{\cos^2\Big[L\sqrt{\frac{\delta\pm|\varepsilon|\sqrt{1-v^2\cos^2\phi}}{\alpha}}\Big]+\sin^2\Big[L\sqrt{\frac{\delta\pm|\varepsilon|\sqrt{1-v^2\cos^2\phi}}{\alpha}}\Big]\frac{(1-\sin\theta\sin\phi)^2}{\cos^2\theta\cos^2\phi}},
\end{eqnarray}
\end{widetext}
We now explore the behavior of different thermoelectric parameters as one alters the relative strength of the hopping parameters $t_1$ and $t_2$. We will be exploring the different transport properties by essentially varying the gap parameter $\delta$ from zero to some finite positive number keeping in mind that we are still in the low energy limit in the vicinity of the $M$-point. This assumption ensures the validity of the Hamiltonian written in Eq. (\ref{Hamil}). The variation of the gap parameter implies a variation in the relative strength between the hopping parameters, $t_1$ and $t_2$. As discussed before, the condition $t_2=2t_1$ corresponds to the gap parameter $\delta=0$ and that $\delta>0$ implies $t_2<2t_1$. For $\delta<0$, the dispersion as depicted in Fig. \ref{Fig1} (h \& i) will not
lead to significant results in the thermoelectric coefficients since there will be no von-Hove singularity in the density of states, as in the case of $\delta>0$.
\begin{figure}[t]
	\includegraphics[width=80mm,height=50.5mm]{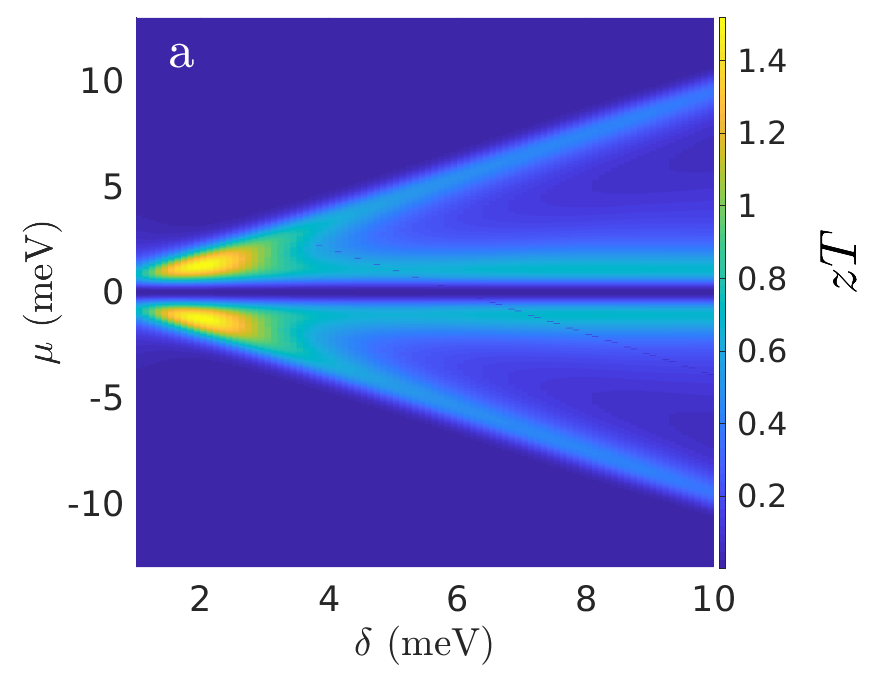}	\includegraphics[width=88mm,height=45.5mm]{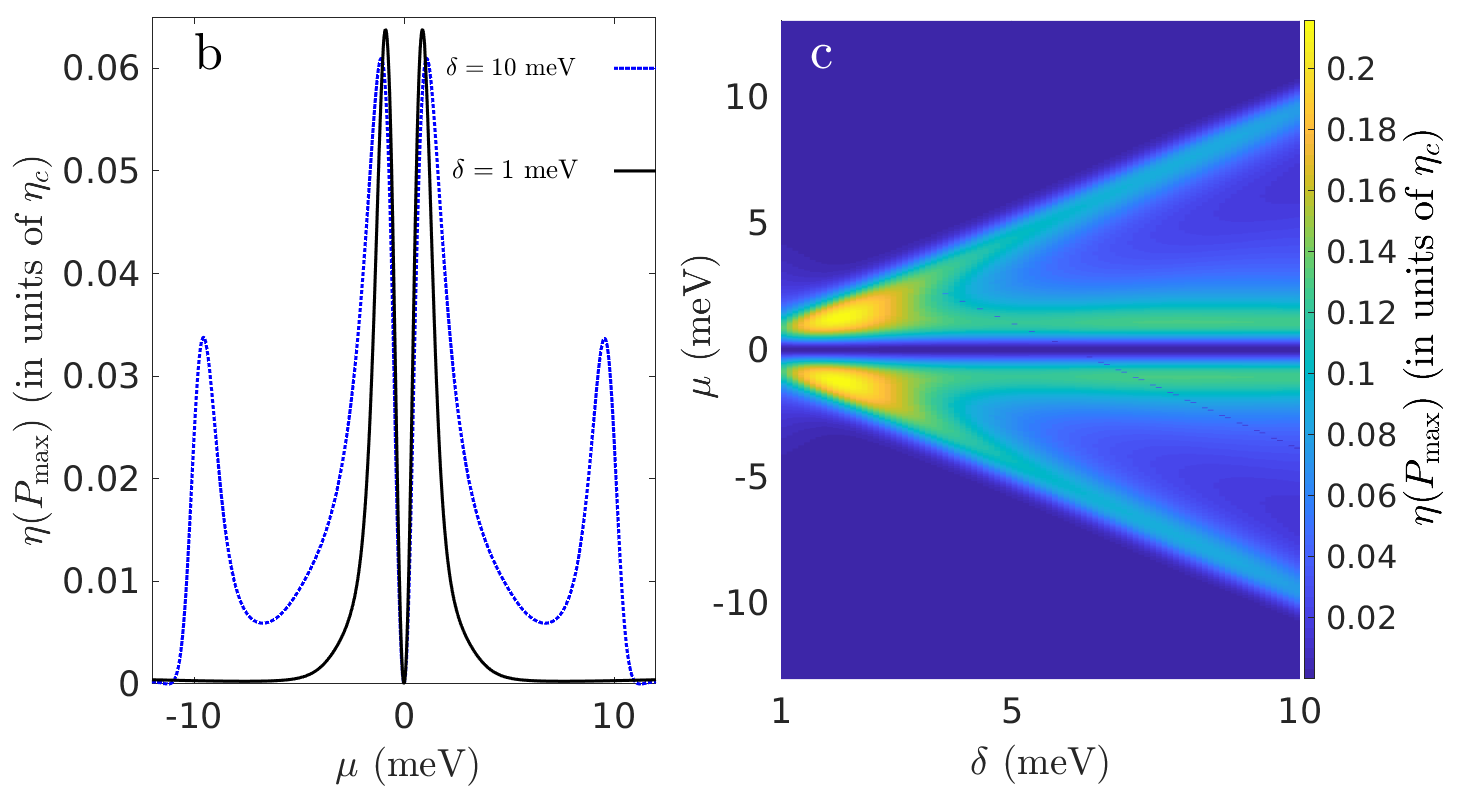} 	\caption{(a) Color plot of $zT$-value as a function of the gap parameter and the chemical potential. (b) Plots of the efficiency at maximum power for $\delta=10$ meV (dotted blue) and $\delta=1$ meV (solid black). (c) Color plot of the efficiency at maximum power as a function of chemical potential and the gap parameter.}
	\label{Fig4}
\end{figure}
\\
\indent
Firstly, we investigate the Seebeck coefficient $S={\mathcal{L}^1}/{ eT \mathcal{L}^0}$ obtained by setting the electronic current under a temperature gradient $(j)$ to zero.
We also look for a condition to derive maximum power output from the system, $
P_{\rm max}=\frac{1}{2}S^2\mathcal{L}^0(\Delta T)^2
$. For optimal operation of the thermoelectric device, its operation with the highest possible efficiency is also to be considered. 
The efficiency of the thermoelectric system is taken to be the ratio of the power to the thermal current. Since the main objective of the paper is to look into the operation of the system at maximum power, we, therefore, stress on the efficiency calculated at the maximum power which is given by \cite{etamax}
\begin{eqnarray}
\eta(P_{\rm max})=\frac{\eta_c}{2}\frac{zT}{2+zT},
\end{eqnarray}
where $\eta_c=\Delta T/T$ is the Carnot efficiency. The system thermal conductivity $\kappa$ appearing in the figure of merit $zT$ is given by $\kappa=(\mathcal{L}^0\mathcal{L}^2-{\mathcal{L}^1}^2)/e^2T\mathcal{L}^0$ which basically is related to the variance of the quantity $\varepsilon-\mu$. It must be noted that the quantum effect should make the ratio $\kappa/T\mathcal{L}_0$ differ from the usual Lorentz number thus ensuring a small decoupling of electronic and thermal currents.
\begin{figure}[t]
	\includegraphics[width=75mm,height=53.5mm]{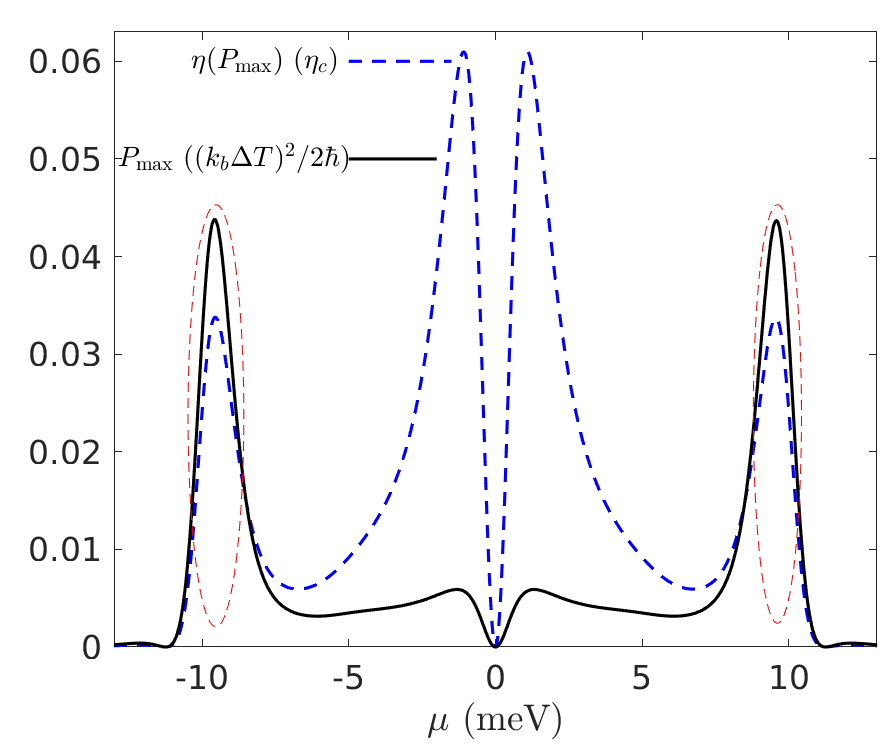}
	\caption{Plots of maximum power (solid black) and the efficiency at maximum power (dashed blue) for the gap parameter set at $\delta=10$ meV. The region enclosed by the ellipse represents the ideal regime of operation of the thermoelectric device.}
	\label{Fig5}
\end{figure}
\\
\indent
For our numerical analysis, we have chosen the parameter $\alpha=0.0075$ eV nm$^2$, $v=0.065$ eV nm and maintaining the temperature $T=5$ K. The dimension of the semi-Dirac system is taken as $(L,W)=(40,20)$ nm. We start by analysing the behaviour of the Seebeck coefficient as a function of the chemical potential and the gap parameter as given in Fig. (\ref{Fig3}). In the subplots (a \& b) of Fig. (\ref{Fig3}), we show the variation of the Seebeck coefficients. The Seebeck coefficient exhibits a sign change when the chemical potential $\mu$ crosses the gap parameter $\delta$ apart from its enhancement around the point $\mu=\delta$, as seen from Fig. \ref{Fig3} (a).
\\
\indent
We then consider the maximum power that the system can generate. Interestingly, the maximum power is boosted whenever the chemical potential matches the gap parameter as can be seen in Fig. \ref{Fig3} (c \& d). 
In order to get an overview of the Seebeck coefficient and maximum power as a function of the chemical potential and the gap parameter, we plot in Fig. (\ref{Fig3} (b \& d)) a color gradient of the Seebeck coefficient and maximum power in the ($\mu$, $\delta$)-plane. Here, the gap parameter is varied from $\delta=1$ meV to $\delta=10$ meV. Note that this variation of a gap parameter corresponds to varying the ratio of hopping strengths from $2t_1<t_2$ to $2t_1=t_2$. 
\\
\indent
As can be observed from Fig. \ref{Fig4} (a), the figure of merit, $zT$-value approaches to about $1.5$ for the chemical potential close to zero. There are also peaks in the $zT$-value whenever $\mu=|\delta|$ whose values range from $0.5-1$. The efficiency when the system operates at maximum power follows the same trend as that of the $zT$-value. Thus, the best efficiency is obtained whenever the chemical potential matches the gap parameter and also is higher for $\mu$ close to zero Fig. \ref{Fig4} (b). The two maximas of the efficiency close to $\mu=0$ as seen in Fig. (\ref{Fig4} (b \& c) \& \ref{Fig5}) are irreleveant since the power factor in these regime is well below the maximum value at $|\mu|=\delta$. To the rescue, the other two maximas of the efficiency at maximum power are obtained whenever the chemical potential matches the gap parameter (Fig. \ref{Fig5}) where the system exhibits the largest maximum power. Thus, it is safe to conclude that the gap parameter in semi-Dirac materials induces the maximum power factor with the best possible efficiency in the thermoelectric nanosystem. The regime for ideal operation of the system is indicated using two ellipses in Fig. \ref{Fig5} where there is a well-balanced maximum of the power factor and efficiency. 
\\
\indent
In summary, we have seen that the Dirac cones in $K$- and $K^\prime$- points merge to a common $M$-point in semi-Dirac materials by altering the relative strength between the two different nearest-neighbor hopping energies, $t_1$ and $t_2$. We have seen that a gap parameter can fully describe the entire process of the merging of Dirac cones in semi-Dirac materials. 
On examining the effect on the thermoelectric properties, we first note that the Seebeck coefficient changes its sign when the Fermi energy crosses the gap along with a corresponding boost in the power factor at $\mu=\delta$. The gap parameter results in the enhancement of the $zT$-value and consequently the efficiency at maximum power. Also, there is a co-maximization of the power factor and efficiency since the highest maximum power obtained is also accompanied by the best possible efficiency one can achieve in the semi-Dirac based thermoelectric system. In short, this paper suggests the possibility of designing an electronic and thermal current decoupled semi-Dirac based nanodevice that can be operated to yield maximum possible power with the best possible efficiency.

This work is an outcome of
the Research and Development work undertaken in the project
under the Visvesvaraya Ph.D. Scheme of Ministry of Electronics and Information Technology, Government of India, being
implemented by Digital India Corporation (formerly Media
Lab Asia). This work was also supported by the Science and
Engineering Research Board (SERB) of the Government of
India under Grant No. EMR/2017/002853.

\end{document}